RF-quantum capacitance of the topological insulator $Bi_2Se_3$ in the bulk depleted regime for field-effect transistors

A. Inhofer[1], J. Duffy[2,3], M. Boukhicha[1], E. Bocquillon[1], J. Palomo[2], K. Watanabe[4], T. Taniguchi[4], I. Estève[5], J.M. Berroir[1], G. Fève[1], B. Plaçais[1], B.A. Assaf[2]

[1] *Laboratoire Pierre Aigrain, Département de Physique, Ecole Normale Supérieure, PSL Research University, Université Paris Diderot, Sorbonne Paris Cité, Sorbonne Université, CNRS, 24 rue Lhomond, 75005, Paris, France*

[2] *Département de Physique, Ecole Normale Supérieure, PSL Research University, CNRS, 24 rue Lhomond, 75005 Paris, France*

[3] *Chemical Engineering Department, Northeastern University, 360 Huntington Avenue, Boston MA, 02115, USA*

[4] *National Institute for Materials Science, 1-1 Namiki, Tsukuba 305-0044, Japan.*

[5] *Institut de Minéralogie, de Physique des Matériaux et de Cosmochimie, UMR 7590 CNRS UPMC-IRD-MNHN, Campus Jussieu, 4 Place Jussieu, 75005 Paris, France.*

**A Metal-dielectric-topological insulator capacitor device based on hBN-encapsulated CVD grown $Bi_2Se_3$ is realized and investigated in the radio frequency regime. The RF quantum capacitance and device resistance are extracted for frequencies as a high as 10 GHz, and studied as a function of the applied gate voltage. The superior quality hBN gate dielectric combined with the optimized transport characteristics of CVD grown $Bi_2Se_3$ ($n\sim 10^{18}cm^{-3}$ in 8 nm) on hBN allow us to attain a bulk depleted regime by dielectric gating. A quantum capacitance minimum and a linear variation of the capacitance with the chemical potential are observed revealing a Dirac regime. The topological surface state in proximity to the gate is seen to reach charge neutrality, but the bottom surface state remains charged and capacitively coupled to the top via the insulating bulk. Our work paves the way towards implementation of topological materials in RF devices.**

**Introduction**

Topological phases of matter have emerged as a fundamental paradigm in the study of condensed matter physics. [1] [2] [3] [4] Topological insulators (TI) are essentially material realizations that stem from this new theoretical paradigm. [4] [5] [6] [7] [8] They are interesting from both the fundamental and applied perspective. Typically, a topological insulator is a material that has an inverted orbital band ordering in the 3D bulk, which leads to the existence of Dirac cones at the surface of the material, at symmetric points in the Brillouin zone. [4] [5] These Dirac fermions are spin-momentum locked and highly robust to backscattering. From the fundamental perspective, a number of novel states of matter have so far been realized in topological insulators. The quantum anomalous Hall state, [9] [10] the Majorana fermion [11] and the quantized Faraday and Kerr effects [12] [13] are examples of such realizations. From the applied perspective, implementation in spintronic data storage devices and high frequency transistors are envisaged. Highly efficient spin-torque switching and spin injection have already been demonstrated in ferromagnet-TI bilayer devices thus establishing potential use in data storage, [14] [15] [16] however, studies aimed at realizing high frequency transistors still lack.

An important step to realize high frequency transistors is to characterize the capacitive response of the TI at radio frequency (RF) and establish its Dirac-like nature. The RF transport regime has already been significantly studied in graphene [17] [18] [19] [20] and more recently in HgTe TIs. [21] [22] In this regime, one can simultaneously measure the quantum capacitance of the Dirac states and the conductivity of the material. Contrary to the constant capacitance-voltage characteristic typical of metal-insulator-metal capacitors (Fig. 1(a)), in metal-insulator-graphene and metal-insulator-topological insulator capacitors (MITI-CAP), the capacitance is a function of the applied voltage (Fig 1(b)). The quantum capacitance being related to the compressibility or the density of states, then allows one to directly measure those quantities. [17] This is, however, only possible for materials that have a low carrier density and a good mobility. For this particular reason, RF capacitance studies have remained highly challenging in Bi-based 3D-TIs. In the case of $Bi_2Se_3$ for example, residual bulk n-doping renders a reliable detection of surface-state signatures difficult. [23] [24] [25] The first step in realizing an RF-transport device based on $Bi_2Se_3$ is a solution to the issue of material quality.

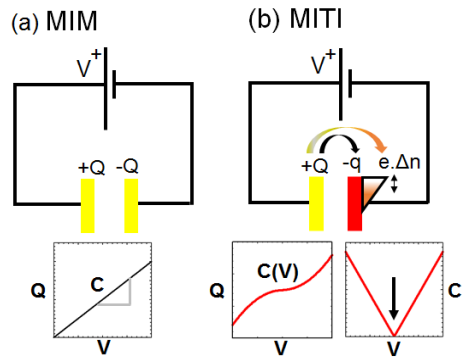

**FIG 1**. (a) Linear charge (Q) vs voltage (V) characteristic of a metal-insulator-metal (MIM) device. The capacitance is constant in this case (b) Non-linear Q vs V curve characteristic of voltage dependent quantum capacitance in a MITI-CAP. The quantum capacitance is due to finite change in density ($e\Delta n$) and chemical potential at the surface of the TI. The arrow in C-V graph indicates the Dirac point.

Motivated by recent positive results on the growth of $Bi_2Se_3$ by chemical vapor deposition (CVD) on mechanically exfoliated hexagonal boron nitride (hBN), [26], [27] we undertake a similar procedure. We first grow $Bi_2Se_3$ by CVD on high-quality hBN, [28] [29] and then transfer a second layer of hBN on top of the grown $Bi_2Se_3$ to realize a capacitor device. The excellent dielectric properties of the hBN used in this work, and the improved transport characteristics of CVD grown $Bi_2Se_3$ on hBN, allow us to observe clear signatures of Dirac surface-states in the RF transport regime. We are able to simultaneously measure both the quantum capacitance and the channel resistance of the device. The capacitance exhibits a linear variation and a minimum versus chemical potential characteristic of Dirac fermions. The resistance shows a strong increase with decreasing voltage in the depleted regime. It does not reach a maximum at the capacitance dip. We argue that this is due to the contributions of the bottom surface in the bipolar regime. Our work provides a first quantitative analysis of the compressibility of $Bi_2Se_3$ in the RF regime and establishes the Dirac nature of the RF response in TIs.

**Growth and characterization**

Bi$_2$Se$_3$ nanoflakes are grown by catalyst free chemical vapor deposition (CVD) using a three zone tube furnace following a procedure similar to what is reported by Xu et al. [26] All growths are performed on high-resistivity Si/SiO$_2$ substrates having an oxide thickness of 300nm, on which we mechanically exfoliate h-BN. The furnace tube is initially pumped down to 8x10$^{-2}$mbar. A powder source of high purity (99.99%) Bi$_2$Se$_3$ is placed in the hot zone (A) of the furnace in a stream of Argon gas (99.999%) flowing at 200 sccm (Fig. 2(a)). The substrate (Fig. 2(b)) is placed downstream from the source in the colder zone (B). Zones (A) and (B) are initially heated up to 300$^0$C in 30 min. (A) is then heated up to 600$^0$C while (B) is only heated up to 400$^0$C in 30 min. These temperatures are maintained for 60 seconds. Both zones are finally cooled down to 200$^0$C in 80 min. An absolute pressure of 3.9 mbar is maintained during the entire process.

An optical microscope image of a characteristic sample is shown in Fig. 2(b,c), before and after the growth respectively. A layer of Bi$_2$Se$_3$ is seen to coat the hBN flakes, but does not nucleate on the SiO$_2$. A Z-contrast scanning electron microscope image shown in Fig. 2(d) confirms nucleation of Bi$_2$Se$_3$ on the hBN flake. The dark spots observed on the light grey flake indicate the presence of heavy atoms such as Bi or Se. This growth mechanism is consistent with previous reports on CVD synthesis of Bi-based TI on hBN. [26] [27]

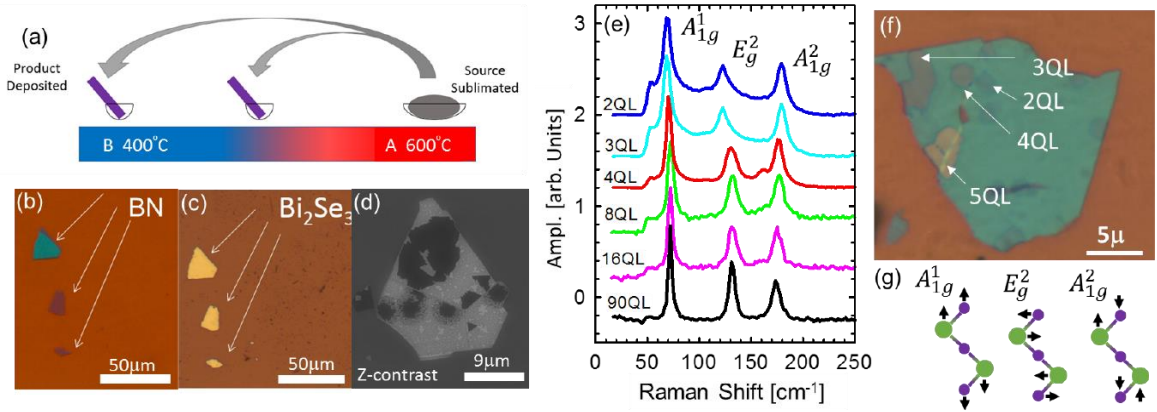

**FIG 2.** (a) Schematic of the CVD growth tube showing the Bi$_2$Se$_3$ source in hot zone A, and substrates in colder zone B, down stream in the Argon flow direction. (b) Optical microscopy image showing hBN exfoliated flakes on SiO$_2$ prior to growth. (c) The same hBN flakes coated with 90QL of Bi$_2$Se$_3$ after the growth. (d) Z-contrast SEM image of Bi$_2$Se$_3$ growth (dark patches) on hBN flake. (e) Microscopic Raman spectroscopy of Bi$_2$Se$_3$ flakes of different thicknesses on hBN. The 2-4QL flakes are shown in (f). Three Raman active peaks are observed in (e) corresponding to the three vibrational modes shown in (g), namely, the $A_{1g}^1$ and $A_{1g}^2$ out-of-plane modes and the $E_g^2$ in-plane mode.

Fig. 2(e) shows Raman spectra obtained using an excitation wavelength of 532nm on Bi$_2$Se$_3$ flakes having thicknesses ranging from 2QL to 90QL. The thinnest flakes studied in Raman spectroscopy (2QL-4QL) all nucleate on the same BN flake, shown in Fig. 2(f). Three characteristic Raman active phonon peaks (Fig. 2(g)) are observed in Fig. 2(e) between 50 and 200 cm$^{-1}$, confirming the presence of a Bi$_2$Se$_3$ layer on the exfoliated hBN. A blueshift of the $A_{1g}^2$ mode and a redshift of the $E_g^2$ and $A_{1g}^1$ modes are observed with decreasing thickness in agreement with previous reports on Raman spectroscopy on $Bi_2Se_3$. [26] [30]

**Device fabrication**

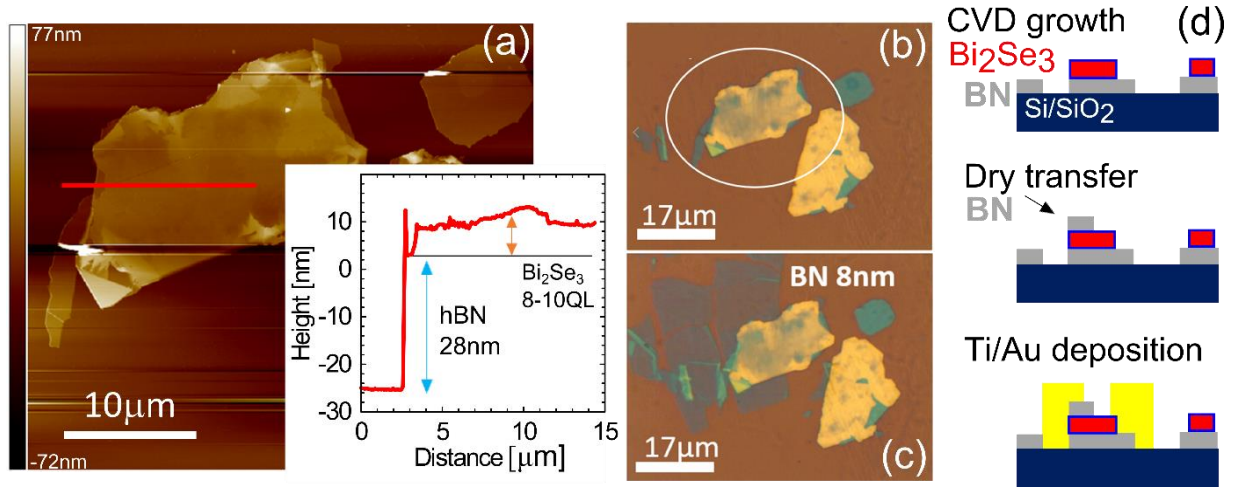

**FIG 3.** (a) AFM image of a selected 8QL thick $Bi_2Se_3$ flake. Inset shows the height as a function of distance measured along the red line shown in the AFM image. The $Bi_2Se_3$ is 8-10QL thick, and the underlying hBN is 28nm thick. (b) Optical microscope image of the flake shown in (a). (c) 8QL $Bi_2Se_3$ flake with an 8nm hBN flake transferred on top using the dry-transfer method. (d) Summary of device processing steps: growth of exfoliated BN, dry transfer of top BN and Ti/Au metal deposition.

Three flakes (6QL, 8QL, 13QL) having different thicknesses obtained from identical growths are selected for device implementation. The 8QL sample is studied in detail. The results from the other three samples are shown in the appendix section. An AFM image of the 8QL sample is shown in Fig. 3(a). We measure the $Bi_2Se_3$ thickness across a line shown in the inset of Fig. 3(a). We find a thickness varying between 8 and 10 quintuple layers (QL). A small peak-to-peak surface roughness of (1-2)QL is detected. A hBN flake is then transferred on top of the grown $Bi_2Se_3$ using the standard dry transfer method, already proven successful for graphene [31] [32] and 2D semiconductors. [33] [34] [35] [36] [37] Optical images of the sample are shown in Fig. 3(b, c) before and after the transfer respectively. The transfer is performed in air, therefore exposing the top surface of $Bi_2Se_3$ to atmosphere. The transferred hBN layer is chosen to be thinner than 10 nm (8 nm in this case), in order to maximize its geometric capacitance and render the quantum capacitance of the $Bi_2Se_3$ experimentally visible. A single e-beam lithography step then allows us to pattern the gate and drain electrodes, as well as a co-planar waveguide. A metallic bilayer of Ti(5 nm)/Au(150 nm) is then deposited. Note that prior to depositing the Ti layer, light argon etching (<10 s) is performed in-situ to remove any native oxides, and to minimize contact resistance. The device processing sequence is summarized in Fig. 3(d). The hBN-encapsulated $Bi_2Se_3$ capacitor device embedded in an RF-waveguide is shown Fig. 4 (a, b).

**RF-transport measurements**

The devices are characterized using radio-frequency (RF) transport measurements for frequencies between 0.03 and 10 GHz using a variable network analyzer, in a cryogenic RF-probe station, as detailed in our previous works. [17] [22] [21] A standard short-open-load-through calibration is performed before the measurement. The S-matrix components are extracted by measuring the reflected and transmitted wave intensity through the device as a function of frequency for different gate voltages using a variable network analyzer. The complex admittance (inverse impedance) is then extracted from the S-matrix components. The real and imaginary parts of device admittance are quantified versus frequency and gate voltage. Proper care is taken to de-embed [38] parasitic capacitive and inductive contributions resulting from the device geometry by measuring a dummy device having identical contact geometry without the $Bi_2Se_3$ flake in between, as well as a conductive through-line. Such measurements also rule out any parasitic RF signals stemming from the substrate $Si/SiO_2$ substrates. All measurements are made at 10K. In what follows, we will focus on measurements made on the 8QL device shown in Fig. 2 and 3.

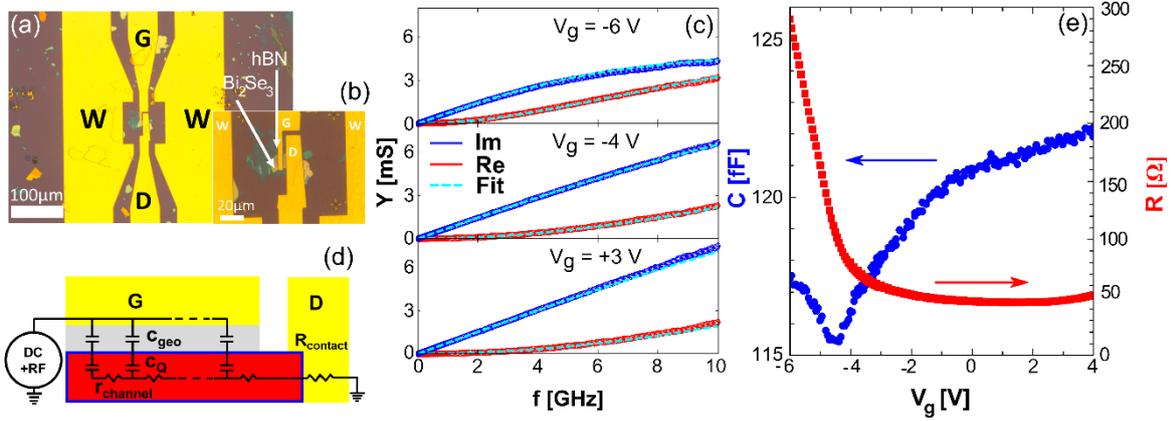

**FIG 4.** (a) Optical microscope image of the finished device with the capacitor shown embedded inside a coplanar waveguide (W). G and D denote the gate and drain contacts respectively. (b) Zoomed in microscope image showing the capacitor device. (c) Real (red) and Imaginary (blue) parts of the RF-admittance as a function of frequency for three typical gate voltages. (d) Schematic of a distributed RC-line model in series with a contact resistance. The capacitance is split into two components the geometric capacitance $C_{geo}$ and the quantum capacitance $C_Q$. (e) Total device capacitance and channel resistance extracted by fitting the model corresponding to the circuit shown in (d) to the data in (c). Curve fits in (c) are shown as dashed lines.

The de-embedded sample admittance versus frequency is shown in Fig. 4 (c) for three different gate voltages. The data is then fit using a distributed RC-line model (of admittance $Y_{RC}$) in series with a contact resistance $R_{contact}$ (Fig. 4(d)), similar to what was previously reported for graphene [17]. The total admittance $Y_{total}$ is given by:

$$Y_{total}(\omega) = \frac{1}{Y_{RC}^{-1}(\omega) + R_{contact}}, \quad (1)$$

Where,

$$Y_{RC} = j\omega C L w \frac{\tanh(L\sqrt{j\omega C/\sigma})}{L\sqrt{j\omega C/\sigma}} \qquad (2)$$

Here, $\omega = 2\pi f$ is the frequency, $j = \sqrt{-1}$ and $L = 4.3\,\mu m$ and $w = 12.8\,\mu m$ are the device length and width, respectively. $\sigma$ is the channel conductivity. $C$ is the total device capacitance:

$$C^{-1} = c_{geo}^{-1} + c_Q^{-1} \qquad (3)$$

Where $c_{geo}$ is the geometric capacitance and $c_Q = e^2 \chi$ is quantum capacitance related to the compressibility $\chi = \frac{\partial n}{\partial \varepsilon_F}$. Here $\varepsilon_F$ represents the chemical potential at the sample surface and $n$ is a 2D carrier density.

The curve fit of $Y_{total}(\omega)$ allows us to separate $C$ and $\sigma$ for different gate voltages. As seen in Fig. 4(c), the model yields an excellent fit to the data up to 10 GHz. Results for $C$ and $R$ from curve fits up to 10 GHz are shown in Fig. 4(e) as a function of gate voltage between -6 V and 1V. The capacitance decreases progressively for decreasing voltage (0 to -5 V), goes through a minimum at about -5 V, then increases again between -5 V and -6 V. With a capacitance dip of 5%, our results agree with previous low frequency capacitance measurements on similar $Bi_2Se_3$ reporting a capacitance dip of 6%. [26]

The resistance exhibits a continuous increase that accelerates near the capacitance minimum. No resistance maximum is observed. A fixed contact resistance $R_c \approx 20\Omega$ was included in the curve fit to the admittance data. Allowing this contact resistance to vary yielded negligible variation compared to the 5-fold increase observed in the channel resistance. We can thus confidently claim, that our two-point RF measurement yields a reliable simultaneous measurement of the quantum capacitance and the channel resistance.

### V. Analysis of the quantum capacitance of top surface states

We next focus on the analysis of quantum capacitance. $c_Q$ can be extracted using Eq. (3), by fixing $c_{geo} = (124 \pm 1)$ fF, the value at which the capacitance is seen to saturate in Fig. 4(e). The measured geometric capacitance is slightly lower than what is expected for hBN having κ=3.2, possibly as a result of the rough surface of $Bi_2Se_3$. The capacitance per unit area $c_Q$ is then determined by dividing the capacitance by the geometrical factors $L$ and $w$. $c_Q$ is shown in Fig. 5 (a). The grey lines in Fig. 5(a) delimit the propagated uncertainty on $c_Q$ due to the uncertainty associated with $c_{geo}$. A powerful consequence of our measurements is the fact that it allows us to determine the local (top surface) chemical potential $\varepsilon_f$ directly from experimental data via the Berglund integral, without a priori knowledge of the band structure. The Berglund integral is written as [39]:

$$\varepsilon_f = \int_0^{V_g} dV \left(1 - \frac{C(V)}{c_{geo}}\right) \quad (4)$$

We use Eq. (4) to extract $\varepsilon_f$ which we now plot as a function of $V_g$ in Fig. 5(b) along with the propagated uncertainty associated with it. We can now plot the quantum capacitance $c_Q$ versus the local chemical potential $\varepsilon_f$. This is shown in Fig. 5(c). The minimum in $c_Q$ defines the chemical potential origin, as it is associated with the position of the Dirac point; and allows us to determine the Fermi energy at zero applied potential to be close to (200±40) meV above the Dirac point. This Fermi energy corresponds to a surface Dirac carrier density of $(3\pm1)\times10^{12}$cm$^{-2}$. Assuming the Dirac point occurs at 200meV below the bottom-most bulk conduction band, as seen in ARPES, [40] [23] we get a Fermi level position of at most 40 meV above conduction band bottom of Bi$_2$Se$_3$ (assumed parabolic with *m\*=0.14m$_0$*). [41] This Fermi-level position yields a slight native bulk n-doping of $2\times10^{18}$cm$^{-3}$. This is a significant improvement compared to pristine quality Bi$_2$Se$_3$ where typically $n > 10^{19} cm^{-3}$ [42] [43] [44] [45] [46] [47] and agrees with previous reports on improved quality samples (see table I). [48] [49]

| At $V_g$=0 | $n_{sheet}$ [cm$^{-3}$] | Thickness QL | Type of samples |
|---|---|---|---|
| This work | **2×10$^{18}$** | 8 | CVD |
| Ref. [45] | 4×10$^{19}$ | 80 | CVD |
| Ref. [48] | 4×10$^{19}$ | 100 | CVD |
| Ref. [49] | **(1-3)×10$^{18}$** | 10 | MBE |
| Ref. [50] | (5-10)×10$^{18}$ | 10 | MBE |
| Ref. [46] | (1-2)×10$^{19}$ | 10 | MBE |
| Ref. [47] | 7×10$^{19}$ | 10 | MBE |
| Ref. [44] | 2×10$^{19}$ | 20 | MBE |
| Ref. [42] | 2×10$^{19}$ | 8 | MBE |
| Ref. [51] | >10$^{19}$ | 10QL pristine | Exfoliated |

**Table I.** Comparison of the Bi$_2$Se$_3$ sample studied in this work to those previously reported.

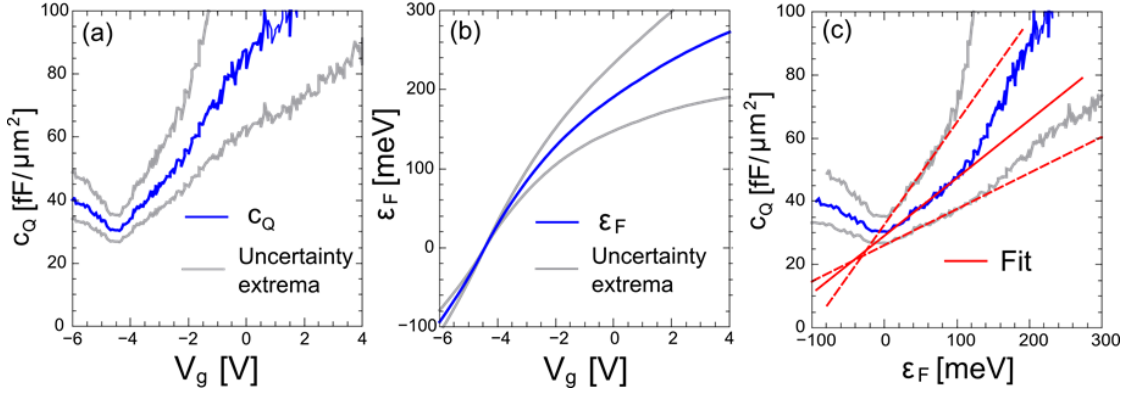

**FIG 5.** (a) Quantum capacitance versus gate voltage at 10K. (b) Surface chemical potential (Fermi energy) versus gate voltage. (c) Quantum capacitance versus Fermi energy. Blue line represents data, grey lines show uncertainty extrema, red lines are linear fits using Eq. 5.

We can additionally perform a linear fit to the $c_Q$ versus $\varepsilon_f$ curve. This is shown as the solid red line, fit to the data in Fig. 5(c). The dashed lines are fit extrema to the edges of the shaded area, and allow us to determine the uncertainty on the extracted slope. The slope allows us to extract the Dirac velocity of the topological surface states, near the Dirac point, from the compressibility χ, which is related to $c_Q$:

$$c_Q = e^2 \chi = \frac{e^2 \varepsilon_f}{2\pi(\hbar v_f)^2} \quad (5)$$

We find a Dirac velocity equal to $(5.8\pm1.4)\times10^5$m/s. This agrees within experimental uncertainty with previously reported velocities that vary between $5\times10^5$m/s and $4.5\times10^5$m/s. [40] [52] [25] It is worthwhile highlighting that while we can perfectly account for variation of the quantum capacitance versus Fermi energy shown in Fig. 5(c), this analysis cannot account for the finite $c_Q$ offset observed at the minimum. From Eq. (5) it is evident that $c_Q$ should go to zero at $\varepsilon_f = 0$. Although this has never been observed experimentally, the typical capacitance offset at the Dirac point observed in graphene, for example, is an order of magnitude lower than what is observed here. [53] Our data suggests the presence of additional capacitive contributions in parallel with that of the top Dirac surface states. A thorough understanding of these capacitive contributions from other transport channels must thus be provided.

## VI. The origin of the capacitance offset: capacitive coupling of top and bottom surface channels

First, it is simple to rule out reminiscent bulk carriers as the source of this offset, simply by computing the screening length for $Bi_2Se_3$ using the measured quantum capacitance. In order to extract the screening length from experimental data, we need to develop and expression that relates the screening length to the 2D density-of-states and to the quantum capacitance. This is discussed in detail in appendix A. The end result is a screening length that scales linearly with the inverse of the 2D quantum capacitance of bulk states $c_Q^{bulk}$:

$$\lambda = \frac{\kappa \varepsilon_0}{c_Q^{bulk}} \quad (6)$$

Note that here $c_Q^{bulk}$ is a quantum capacitance, related to a surface charge accumulation or depletion that screens the electric field over a finite length. $\kappa \approx 100$ is the generally accepted static dielectric constant of Bi$_2$Se$_3$. [51] [54] [55] [56] [50] $\varepsilon_0 = 8.85 \times 10^{-12} F/m$ is the permittivity of free space. If we assume that all measured quantum capacitance contributions are due to screening, then for $c_Q^{bulk} < 100$ fF/μm², $\lambda$ exceeds 10 nm and the thickness of the sample (8nm); implying that full bulk depletion is possible. Near the capacitance minimum, it is thus highly unlikely that reminiscent charge carriers from the bulk contribute to the transport.

Consequently, we can consider the situation of a fully depleted insulating bulk that contributes a geometric capacitance, between two metallic surfaces. The Dirac character of the top metallic surface is proven by the quantum capacitance observed in Fig. 5. However, the character of the bottom surface is not evident in our experiments. Experimentally, we can still determine the net quantum capacitance contribution of this bottom metallic layer, assuming it couples capacitively in parallel with the top surface via the insulating bulk:

$$c_Q = c_Q^{TTSS}(\varepsilon_f) + \left(\frac{1}{c_g^{Bulk}} + \frac{1}{c_Q^{Bottom}(\varepsilon_f)}\right)^{-1} \quad (7)$$

Here, $\varepsilon_f$ denotes the chemical potential of the top surface, $c_Q^{TTSS}$ is top surface quantum capacitance and $c_Q^{Bottom}$ is that of the bottom metallic surface. The geometric capacitance of the insulating bulk with a thickness $d_{Bi2Se3} \approx 8\ nm$ is given by:

$$c_g^{Bulk} = \frac{\kappa \varepsilon_0}{d_{Bi2Se3}} \approx 110 \text{ fF/μm}^2 \quad (8)$$

When $c_Q^{TTSS}(\varepsilon_f = 0)=0$, an offset equal to $(30 \pm 5)$ fF/μm² has to result from the other term in Eq. 7). Eq. (7) and Eq. (8) yield $c_Q^{Bottom} = (40 \pm 10)$fF/μm². It can be easily shown that the quantum capacitance expected from a quadratically dispersing Bi$_2$Se$_3$ interfacial 2DEG significantly exceeds this value:

$$\frac{h^2 \pi}{e^2 m^*} = \frac{94\text{fF}}{\text{μm}^2} > c_Q^{Bottom}$$

using the effective mass of Bi$_2$Se$_3$, $m^* = 0.14 m_0$. [41] This is therefore an unlikely scenario.

The last scenario to consider is that the bottom Dirac TSS having a finite Fermi energy at the charge neutrality point of the top TSS, yields the offset in the quantum capacitance. Using Eq. (5), we estimate, for a chemical potential close to (170±40)meV above the Dirac point (with $v_f \approx 5 \times 10^5 m/s$), a Dirac quantum capacitance $c_Q^{Bottom} = (40 \pm 10)$fF/μm². We conclude that the most likely explanation to the observed offset is presence of the bottom TSS that couples capacitively to the top TSS via the insulating bulk.

## VII. Charging curve of a bulk-depleted topological insulator

In order to get further insight on the charging mechanism expected in such as situation, we develop in Fig. 6(a,b) a model that describe the charging of both the top and bottom TSS coupled via the bulk, as a function of a top gate voltage. This capacitance model is summarized by the circuit shown in Fig. 6(b). The total quantum capacitance corresponding to this circuit is given by Eq. (7).

The quantum capacitance of the top TSS $c_Q^{TTSS}$ is computed using Eq. (5), with $v_f \approx 5 \times 10^5 m/s$. [25] Note that in Eq. (7) $\varepsilon_f$ is the top surface chemical potential. Hence, the expression for that of the bottom TSS $c_Q^{Bottom}(\varepsilon_f)$ is not simply given by Eq. (5). We show in the appendix, that the effective capacitance of the bottom TSS that is capacitively coupled via the insulating bulk is given by:

$$\left(\frac{1}{c_g^{Bulk}} + \frac{1}{c_Q^{BTSS}(\varepsilon_f)}\right)^{-1} = c_g^{Bulk}\left[1 - \frac{1}{\left(\sqrt{1 + \frac{e^2|\varepsilon_F + W|}{\pi(\hbar v_f)^2 c_g^{Bulk}}}\right)}\right] \quad (9)$$

$c_Q^{BTSS}(\varepsilon_f)$ is the bottom TSS Dirac quantum capacitance as a function of the top surface chemical potential. W is the band offset between the top and bottom TSS. It is the only adjustable parameter in the model. The W term allows the top and bottom TSS to have a different local chemical potential (Fig. 6(c)), as has been reported in previous studies. [57] [58]

The contribution of $c_Q^{TTSS}$ is shown in red in Fig. 6(a). The evolution of the effective capacitance of the bottom TSS $c_Q^{BTSS}$ in series with the insulating bulk as a function of top-surface chemical potential $\varepsilon_f$ is shown in dashed magenta. For W ≈ 140 meV, the Eq. (9) is in good agreement with the data as seen in Fig. 6(a). The quantum capacitance associated to the bottom surface is seen to flatten out at large values of $\varepsilon_f$ due to an enhanced screening of the electric field at large charge carrier density by the top-TSS. Note that this treatment is only valid when the bulk is fully depleted (region (I) in Fig. 6(a)); the quantum capacitance from populated bulk bands should be taken into account at larger Fermi energies (region II in Fig. 6(a)), however our data is not precise enough to provide any quantitative analysis in this region.

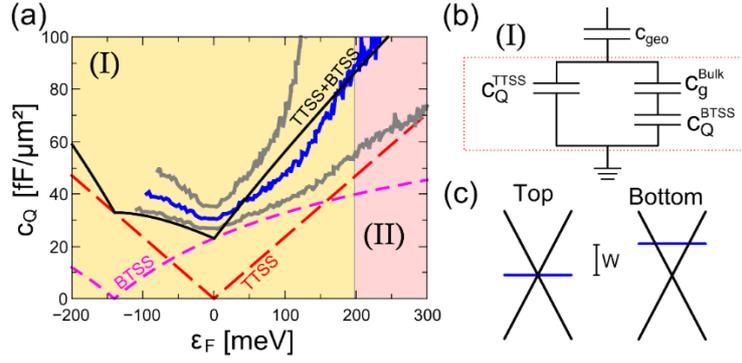

**FIG 6.** (a) Quantum capacitance versus surface chemical potential data (blue) compared to model of topological insulator shown in (b). Gray curves represent upper and lower bounds of experimental data, black solid is the total quantum capacitance resulting from this model, the dashed red line is that of the top TSS, the dashed magenta line is that of the bottom TSS. (b) MITI-CAP model of a topological insulator in the bulk depleted regime where the top TSS ($c_Q^{TTSS}$) is capacitively coupled to the bottom TSS ($c_Q^{BTSS}$) via the insulating bulk assumed to have a geometrical capacitance ($c_g^{Bulk}$). (c) Top and bottom Dirac surface states with respective chemical potential. W is the chemical potential offset The bottom TSS remains filled when the top TSS is at the charge neutrality.

We have thus shown that the measured quantum capacitance can be reliably explained by a model of two capacitively coupled Dirac surface states. Such a strong bottom-TSS contribution to the compressibility has been disregarded in HgTe samples studied previously. [21] [59] It becomes important here due to the large permittivity, the large band gap and the nanometric thickness of the $Bi_2Se_3$ crystals grown by CVD. Importantly, for a top-surface chemical potentials $\varepsilon_F$ between 0 and about -140 meV, the top and bottom surface have different carrier polarity, i.e. p-type and n-type, respectively.

## VIII. THE BEHVIOR OF THE RESISTANCE CHANNEL

Next, we qualitatively discuss the variation of the channel resistance. Near the capacitance minimum, a fast increase of the channel resistance is observed, likely corresponding to the depletion of surface carriers on the top and bottom surfaces, but no resistance maximum is reached. Since the Fermi energy is likely higher above the Dirac point at the bottom surface, bottom Dirac electrons provide a parallel conduction channel that has a much lower resistivity. The effective resistance from both top and bottom parallel channels is hence dominated by the resistivity of the bottom channel. Therefore, one does not expect to observe an ambipolar resistance maximum as long as the bottom Dirac cone remains heavily occupied.

Lastly, going up in Fermi energy away from the capacitance minimum, the resistance curve flattens out and eventually increases at large positive values of $V_g$ (Fig. 4(e)). We point out, that this increased scattering at high charge carrier density in top TSS is likely associated to sub-band scattering as reported in Ref. [22] [21]. It is worth noting that any issue with the contact resistance can be ruled out. First, the $Au/Bi_2Se_3$ interface is not under the gate stack and therefore remains ohmic throughout the entire experiment, since Bi$_2$Se$_3$ remains n-doped away from the gate stack. Second, even when the top surface is very close to neutrality, the bottom surface of the sample remains carrier doped, thus ensuring a good connection with the source contact. This highlights the strength of our measurement and the local nature of the quantum capacitance in our experiment.

**Conclusion**

In conclusion, we have realized implementation of CVD grown Bi$_2$Se$_3$ in RF capacitor devices and reported the observation of the quantum capacitance of the top Dirac surface state and its variation versus gate voltage. The reduced electron doping of Bi$_2$Se$_3$ grown in these conditions, and the use of a high quality hBN dielectric allows us to quantify the quantum capacitance of Bi$_2$Se$_3$ and observe its minimum resulting from the top topological surface Dirac point. A detailed analysis of the field effect mechanism in thin Bi$_2$Se$_3$ flakes shows that a bulk depleted regime can be reached at an accessible gate voltage in hBN-encapsulated Bi$_2$Se$_3$ allowing to investigate Dirac physics. We have lastly modeled the capacitance-voltage curve of a TI slab consisting of two surface states separated by an insulating bulk, and confirmed its correspondence with our data. As an outlook, a dual-gated device might allow to electrostatically compensate the chemical doping asymmetry between the two surfaces. [57] Topological materials with even larger dielectric constants will also be interesting to investigate. [60] [61] Overall, our work establishes hBN encapsulated Bi$_2$Se$_3$ as a promising platform to motivate future work on implementation in high frequency transistors.

**Appendix A. Quantum capacitance and dielectric screening of electron-doped 3D semiconductor**

The quantum capacitance is directly related to the compressibility – the variation of the 2D electrical charge density per unit chemical potential $\frac{\partial n_{2D}}{\partial \mu_T}$:

$$c_Q = e^2 \frac{\partial n_{2D}}{\partial \mu_T} = e^2 \frac{\partial}{\partial \mu_T}\left[\int_0^d \frac{\rho(z)}{e} dz\right] \quad (A1)$$

Here n$_{2D}$ is the surface carrier density, $\mu_T \equiv \varepsilon_F$ is the top surface chemical potential and $\rho(z)$ the 3D charge density. The *z-axis* is chosen to be perpendicular to the surface of the material and therefore parallel to the electric field. The top BN-Bi$_2$Se$_3$ interface is at *z=0*.

From the conservation of electrochemical potential we get: $\mu_T = -eV_0$.

$V_0 = V(z = 0)$ is the electric potential at the dielectric-semiconductor interface. The quantum capacitance can then be written as:

$$c_Q = e^2 \frac{\partial n_{2D}}{\partial \mu_T} = -\frac{\partial}{\partial V_0}\left[\int_0^\infty \rho(z)dz\right]$$

Poisson's law relates the charge to the electric potential:

$$\rho(z) = \kappa\varepsilon_0 \frac{\partial^2 V}{\partial z^2}$$

$\kappa$ is the dielectric constant and $\varepsilon_0 = 8.85 \times 10^{-12} F/m$ is the permittivity of free space.

Plugging $\rho(z)$ into $c_Q$ and performing the integration gives:

$$c_Q = -\kappa\varepsilon_0 \frac{\partial}{\partial V_0}\left(\left[\frac{\partial V}{\partial z}\right]_{z\to\infty} - \left[\frac{\partial V}{\partial z}\right]_{z=0}\right)$$

We assume $\left[\frac{\partial V}{\partial z}\right]_{d\to\infty} = -E(d \to \infty) = 0$ (infinite slab)

Then,

$$c_Q = \kappa\varepsilon_0 \frac{\partial}{\partial V_0}\left[\frac{\partial V}{\partial z}\right]_{z=0} \qquad (A2)$$

Hence, the quantum capacitance measures the changing local (top surface) depletion/accumulation profile $\left[\frac{\partial V}{\partial z}\right]_{z=0}$ induced by the applied surface electrical potential $V_0$. Given in this form, the quantum capacitance from a 3D semiconductor is conceptually simple to understand but not straight forward to quantify experimentally. We will next relate this quantum capacitance to the screening length.

Using the Poisson equation again we can show that since:

$$\frac{\rho(z)}{\kappa\varepsilon_0} = \frac{\partial^2 V}{\partial z^2}$$

We can write:

$$\int_0^\infty \frac{\rho(z)}{\kappa\varepsilon_0}\frac{\partial V}{\partial z}dz = \int_0^\infty \frac{\partial^2 V}{\partial z^2}\frac{\partial V}{\partial z}dz$$

$$\int_{V(0)}^0 \frac{\rho(V)\partial V}{\kappa\varepsilon_0} = -\frac{1}{2}\left(\frac{\partial V}{\partial z}\right)^2_{z=0}$$

Again assuming V and $\frac{\partial V}{\partial z}$ tend to 0 at ∞, and we get:

Or,

$$\sqrt{2\int_0^{V(z=0)} \frac{\rho(V)\partial V}{\kappa\varepsilon_0}} = \left[\frac{\partial V}{\partial z}\right]_{z=0}$$

With $V(z=0)$ by definition equal to $V_0$.

Then, the term $\frac{\partial}{\partial V_0}\left[\frac{\partial V}{\partial z}\right]_{z=0}$ in the quantum capacitance can be written as:

$$\frac{\partial}{\partial V_0}\left[\frac{\partial V}{\partial z}\right]_{z=0} = \frac{\partial}{\partial V_0}\left[\sqrt{2\int_0^{V_0}\frac{\rho(V)\partial V}{\kappa\varepsilon_0}}\right] = \frac{\frac{\partial}{\partial V_0}\int_0^{V_0}\rho(V)\partial V}{\sqrt{2\kappa\varepsilon_0\int_0^{V_0}\rho(V)\partial V}}$$

$$\frac{\partial}{\partial V_0}\left[\frac{\partial V}{\partial z}\right]_{z=0} = \frac{\rho(V_0)}{\sqrt{2\kappa\varepsilon_0\int_{V_0}^0\rho(V)\partial V}}$$

Or,

$$\frac{\partial}{\partial V_0}\left[\frac{\partial V}{\partial z}\right]_{z=0} = \frac{\rho(V_0)}{\kappa\varepsilon_0\left[\frac{\partial V}{\partial z}\right]_{z=0}} = \frac{\rho(V_0)}{\kappa\varepsilon_0 E(0)}$$

Finally, combing this result with Eq. (A2) we get the quantum capacitance:

$$c_Q = \frac{\rho(V_0)}{E(0)} \quad (A3)$$

From Gauss' law, we get:
$$\frac{\partial E}{\partial z} = \frac{\rho}{\kappa \varepsilon_0}$$

Integrating from 0 to infinity and assuming E tends to zero at infinity, yields:

$$\kappa \varepsilon_0 E(0) = -e \int_0^\infty n_0 e^{-\frac{z}{\lambda}} dz = e n_0 \lambda$$

Here, the charge distribution is assumed to follow an exponential decay into the sample:
$$\rho(z) = -e n_0 e^{-\frac{z}{\lambda}},$$
where $n_0$ is the charge at $z=0$ and $\lambda$ is an effective screening length.

We get:
$$\lambda = \frac{\kappa \varepsilon_0 E(0)}{e n_0} = \frac{\kappa \varepsilon_0 E(0)}{\rho(V_0)}$$

According to (A3), we thus have:
$$\lambda = \frac{\kappa \varepsilon_0}{c_Q^{bulk}}$$

Note that here $c_Q^{bulk}$ is a quantum capacitance related a 2D compressibility from a charge accumulation/depletion profile. It is related to the 2D density of states of this charging profile. The screening length $\lambda$ is not the Thomas-Fermi screening length, but rather a screening length parameter that varies with the chemical potential and depends on the 2D quantum capacitance of a surface charging/depletion layer. We have thus derived the dependence of the screening length on the quantum capacitance that we measured experimentally.

**Appendix B. Quantum capacitance from two Dirac surfaces states coupled via an insulating bulk**

A capacitance model accounting for two Dirac surface states coupled via an insulating bulk (Fig. 6(b)) results in the following expression:

$$c_Q(\mu_T) = c_Q^{TTSS}(\mu_T) + \left( \frac{1}{c_g^{Bulk}} + \frac{1}{c_Q^{BTSS}(\mu_T)} \right)^{-1}$$

Here $c_g^{Bulk}$ is the bulk geometric capacitance, and $\mu_T \equiv \varepsilon_F$ is the top surface chemical potential.

The expression for the quantum capacitance of a single Dirac cone at the top surface (Eq. 5 in manuscript) is straight forward to determine:

$$c_Q^{TTSS}(\mu_T) = \frac{e^2 |\mu_T|}{2\pi(\hbar v_f)^2} \quad (B1)$$

$v_f$ is the Fermi velocity.

The expression for the quantum capacitance of single Dirac cone at the bottom surface, is more challenging to extract:

$$c_Q^{BTSS}(\mu_T) = \frac{e^2|\mu_B(\mu_T)|}{2\pi(\hbar v_f)^2}$$

A knowledge of $\mu_B(\mu_T)$ is required. This can be modelled by carefully studying electrochemical equilibrium in a system of two Dirac fluids in parallel, coupled by an insulating capacitive layer.

The electrochemical potential $\mu^*$ imposed by the metallic contact ensures equilibrium and allows us to write:

$$\mu^* = \mu_T - eV_T + W = \mu_B - eV_B \qquad (B2)$$

Here $\mu_{T/B}$ and $V_{T/B}$ are respectively the chemical potential and the electric potential at the top/bottom surface, and $W$ is a work function term that includes contributions that allows band misalignment such as band bending and surface charging due to impurities which essentially lead to a band offset between the top and bottom surface Dirac point.

The 2D carrier density $n_{2D}$ as a function of chemical potential $\mu$ for a non-spin degenerate Dirac surface states can be written as:

$$n = sgn(\mu)\frac{\mu^2}{4\pi(\hbar v_f)^2} \qquad (B3)$$

The charge density is simply given by $\rho = -ne$.

We now apply Gauss' law to find an expression for the charge density at the top and bottom interfaces:

For the bottom surface we have:
$$c_g^{Bulk}(V_B - V_T) = -en_B \qquad (B4)$$

$c_{TG}$ is the geometric top gate capacitance, and $n_{T/B}$ is the carrier density of the top/bottom surface. $V_G$ is the gate voltage. We can now proceed and compute the variation of the quantum capacitance as a function of the chemical potential of the top surface.

By plugging into eq. (B4) the expression for $V_T$ and $V_B$ determined from (B2) and that of $n_T$ and $n_B$ from (B3) we get:

$$c_g^{Bulk}(\mu_B - \mu_T - W) = -e^2 sign(\mu_B)\frac{\mu_B^2}{4\pi(\hbar v_f)^2}$$

We then set up a second degree equation to extract $\mu_B$ as a function of $\mu_T$

$$\frac{4\pi(\hbar v_f)^2}{e^2}c_g^{Bulk}(\mu_B - \mu_T - W) + sign(\mu_B)\mu_B^2 = 0$$

We get four solutions, two of which are inconsistent with the sign of $\mu_B$:
For $\mu_B > 0$:

$$\mu_B = \frac{2\pi(\hbar v_f)^2}{e^2}c_g^{Bulk}\left(-1 \pm \sqrt{1 + \frac{e^2(\mu_T + W)}{\pi(\hbar v_f)^2 c_g^{Bulk}}}\right)$$

Only, the solution with the + sign satisfies $\mu_B > 0$.
Similarly for $\mu_B < 0$, we get only one satisfactory solution:

$$\mu_B = \frac{2\pi(\hbar v_f)^2}{e^2}c_g^{Bulk}\left(1 - \sqrt{1 + \frac{e^2(\mu_T + W)}{\pi(\hbar v_f)^2 c_g^{Bulk}}}\right)$$

Finally, we have $\mu_B$ as a function of $\mu_T$:

$$\mu_B(\mu_T) = \pm \frac{2\pi(\hbar v_f)^2}{e^2} c_g^{Bulk} \left(1 - \sqrt{1 + \frac{e^2(\mu_T + W)}{\pi(\hbar v_f)^2 c_g^{Bulk}}}\right)$$

This expression can then be used to find $c_Q^{BTSS}$:

$$c_Q^{BTSS}(\mu_T) = c_g^{Bulk} \left(-1 + \sqrt{1 + \frac{e^2(\mu_T + W)}{\pi(\hbar v_f)^2 c_g^{Bulk}}}\right) > 0$$

We finally obtain the expression for the total quantum capacitance:

$$c_Q(\mu_T) = c_Q^{TTSS}(\mu_T) + \frac{c_g^{Bulk} c_Q^{BTSS}(\mu_T)}{c_g^{Bulk} + c_Q^{BTSS}(\mu_T)}$$

$$c_Q(\mu_T) = c_Q^{TTSS}(\mu_T) + c_g^{Bulk} \left[1 - \frac{1}{\left(\sqrt{1 + \frac{e^2(\mu_T + W)}{\pi(\hbar v_f)^2 c_g^{Bulk}}}\right)}\right]$$

Finally replacing $\mu_T$ by $\varepsilon_F$, to keep a coherent notation, we get:

$$c_Q(\mu_T) = \frac{e^2|\varepsilon_F|}{2\pi(\hbar v_f)^2} + c_g^{Bulk} \left[1 - \frac{1}{\left(\sqrt{1 + \frac{e^2|\varepsilon_F + W|}{\pi(\hbar v_f)^2 c_g^{Bulk}}}\right)}\right] \quad (B5)$$

We highlight the simplicity of this model from an experimental viewpoint, since most parameters can be determined independently from previous measurements. The only adjustable parameter is the band offset $W$.

## **Appendix C. Thickness dependence**

We have also measured two additional MITI-devices having respectively thinner and thicker $Bi_2Se_3$. The device characteristics are summarized in the Table II. Both devices exhibit a capacitance that varies with voltage. The quantum capacitance is extracted in each case using the analysis described in the manuscript. We did not observe a capacitance minimum in those devices. The quantum capacitance versus gate voltage curve for the 6QL and 14QL samples are shown in Fig. 7(a) and compared to that of the 8QL sample. The experimental error bars are quite large in the 6QL near V=0 since the top hBN used for this sample is quite thick (25nm) and has a small geometrical capacitance. The quantum capacitance measurement is thus less precise in this sample.

This comparison allows us to get an idea about how the capacitance offset depends on thickness. Recall in Eq. (10), the offset is shown to depend on both $c_g^{Bulk}$ and $W$ the Fermi energy offset. $c_g^{Bulk}$ is inversely proportional to the $Bi_2Se_3$ thickness. In Fig. 7(b) we compare the smallest quantum capacitance values measured for each sample. The capacitance minima are plotted versus sample thickness, and compared to the variation of Eq. (10) with sample thickness and $W$. The data suggests that varying thickness yields a changing $c_g^{Bulk}$ and $W$. In the thicker sample, even though the bottom surface is further decoupled from the top gate, its influence on the quantum capacitance is also smaller. This is due to $c_g^{Bulk}$ becoming smaller when the sample thickness increases (see Eq. (10)). This agrees with the fact the capacitance offset observed in HgTe (approximately 70nm) [21] is smaller than that measured in $Bi_2Se_3$ 8QL.

While we cannot draw more conclusions from this analysis, we motive further work on the thickness dependence and the question of capacitive top-bottom coupling.

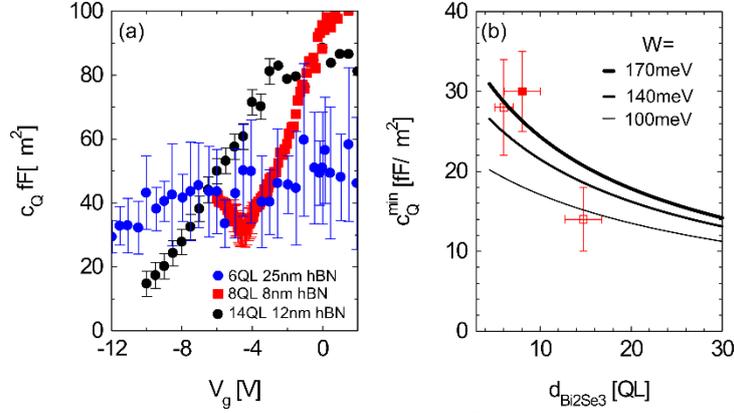

**FIG 7**. (a) Quantum capacitance versus gate voltage for the 3 $Bi_2Se_3$ samples. (b) Smallest $c_Q$ measured in each respective sample. A full square indicates the sample for which we observe the Dirac point feature. An empty square is used for the other. The thickness (horizontal) error bar accounts for sample roughness. Solid lines represent the variation of Eq. (9) versus thickness for different values of W.

|  | $Bi_2Se_3$ thickness [QL] | hBN thickness [nm] |
|---|---|---|
| **Sample 1** | 8QL | 8nm |
| Sample 2 | 14QL | 12nm |
| Sample 3 | 6QL | 25nm |

**Table 2.** Sample list showing corresponding $Bi_2Se_3$ and hBN thickness.

**Acknowledgements.** We acknowledge Michael Rosticher's valuable assistance with cleanroom equipment and the Pascal Morfin for his assistance with growth equipment setup. We also acknowledge discussions with H. Graef. BAA and JD acknowledge funding from ANR-LabEx grant ENS-ICFP ANR-10-LABX-0010/ANR-10-IDEX-0001-02PSL. JD is partially supported by the Northeastern University Coop fellowship. The scanning electron microscope (SEM) facility of the Institut de Minéralogie et de Physique des Milieux Condensés is supported by Région Ile-de-France grant SESAME 2006 N°I-07-593/R, INSU-CNRS, INP-CNRS, University Pierre et Marie Curie – Paris 6, and by the French National Research Agency (ANR) grant no. ANR-07-BLAN-0124-01. Growth of hexagonal boron nitride crystals was supported by the Elemental Strategy Initiative conducted by the MEXT, Japan and JSPS KAKENHI Grant Numbers JP15K21722.

[1]   F. D. M. Haldane, Phys. Rev. Lett. **61**, 2015 (1988).

[2]   Q. Niu, D. J. Thouless, and Y. S. Wu, Phys. Rev. B **31**, 3372 (1985).

[3]   C. L. Kane and E. J. Mele, Phys. Rev. Lett. **95**, 226801 (2005).

[4]   B. A. Bernevig, T. L. Hughes, and S.-C. Zhang, Science (80-. ). **314**, 1757 (2006).

[5]   M. Z. Hasan and C. L. Kane, Rev. Mod. Phys. **82**, 3045 (2010).

[6]   X.-L. Qi and S.-C. Zhang, Rev. Mod. Phys. **83**, 1057 (2011).

[7]   Y. Ando, J. Phys. Soc. Japan **82**, 102001 (2013).

[8]     Y. Ando and L. Fu, Annu. Rev. Condens. Matter Phys. **6**, 361 (2015).

[9]     C.-Z. Chang, W. Zhao, D. Y. Kim, H. Zhang, B. A. Assaf, D. Heiman, S.-C. Zhang, C. Liu, M. H. W. Chan, and J. S. Moodera, Nat Mater **14**, 473 (2015).

[10]    C.-Z. Chang, J. Zhang, X. Feng, J. Shen, Z. Zhang, M. Guo, K. Li, Y. Ou, P. Wei, L.-L. Wang, Z.-Q. Ji, Y. Feng, S. Ji, X. Chen, J. Jia, X. Dai, Z. Fang, S.-C. Zhang, K. He, Y. Wang, L. Lu, X.-C. Ma, and Q.-K. Xue, Science (80-. ). **340**, 167 (2013).

[11]    E. Bocquillon, R. S. Deacon, J. Wiedenmann, P. Leubner, T. M. Klapwijk, C. Brüne, K. Ishibashi, H. Buhmann, and L. W. Molenkamp, Nat. Nanotechnol. **12**, 137 (2017).

[12]    L. Wu, M. Salehi, N. Koirala, J. Moon, S. Oh, and N. P. Armitage, Science (80-. ). **354**, 1124 (2016).

[13]    A. Shuvaev, V. Dziom, Z. D. Kvon, N. N. Mikhailov, and A. Pimenov, Phys. Rev. Lett. **117**, 117401 (2016).

[14]    a. R. Mellnik, J. S. Lee,  a. Richardella, J. L. Grab, P. J. Mintun, M. H. Fischer,  a. Vaezi,  a. Manchon, E. -a. Kim, N. Samarth, and D. C. Ralph, Nature **511**, 449 (2014).

[15]    M. Jamali, J. S. Lee, J. S. Jeong, F. Mahfouzi, Y. Lv, Z. Zhao, B. K. Nikolić, K. A. Mkhoyan, N. Samarth, and J. Wang, Nano Lett. **15**, 7126 (2015).

[16]    J. Han, A. Richardella, S. A. Siddiqui, J. Finley, N. Samarth, and L. Liu, Phys. Rev. Lett. **119**, 77702 (2017).

[17]    E. Pallecchi,  a. C. Betz, J. Chaste, G. Fève, B. Huard, T. Kontos, J.-M. Berroir, and B. Plaçais, Phys. Rev. B **83**, 125408 (2011).

[18]    Q. Wilmart, A. Inhofer, M. Boukhicha, W. Yang, M. Rosticher, P. Morfin, N. Garroum, G. Fève, J.-M. Berroir, and B. Plaçais, Sci. Rep. **6**, 21085 (2016).

[19]    A. C. Betz, S. H. Jhang, E. Pallecchi, R. Ferreira, G. Fève, J.-M. Berroir, and B. Plaçais, Nat. Phys. **9**, 109 (2013).

[20]    Y.-M. Lin, C. Dimitrakopoulos, K. A. Jenkins, D. B. Farmer, H.-Y. Chiu, A. Grill, and P. Avouris, Science (80-. ). **327**, 662 (2010).

[21]    A. Inhofer, S. Tchoumakov, B. A. Assaf, G. Fève, J. M. Berroir, V. Jouffrey, D. Carpentier, M. O. Goerbig, B. Plaçais, K. Bendias, D. M. Mahler, E. Bocquillon, R. Schlereth, C. Brüne, H. Buhmann, and L. W. Molenkamp, Phys. Rev. B **96**, 195104 (2017).

[22]    A. Inhofer, Probing AC Compressibility of 3D HgTe and Bi2Se3 Topological Insulators at High Electric Fields: Evidence for Massive Surface States., PhD Thesis, Ecole Normale Supérieure, (2017).

[23]    J. G. Analytis, J.-H. Chu, Y. Chen, F. Corredor, R. D. McDonald, Z. X. Shen, and I. R. Fisher, Phys. Rev. B **81**, 205407 (2010).

[24]    M. Neupane, A. Richardella, J. Sánchez-Barriga, S. Xu, N. Alidoust, I. Belopolski, C. Liu, G. Bian, D. Zhang, D. Marchenko, A. Varykhalov, O. Rader, M. Leandersson, T. Balasubramanian, T.-R. Chang, H.-T. Jeng, S. Basak, H. Lin, A. Bansil, N. Samarth, and M. Z. Hasan, Nat. Commun. **5**, 3841 (2014).

[25]    K. He, Y. Zhang, K. He, C. Chang, C. Song, L. Wang, X. Chen, J. Jia, Z. Fang, X. Dai, W.

Shan, S. Shen, Q. Niu, X. Qi, S. Zhang, X.-C. Ma, and Q.-K. Xue, Nat. Phys. **6**, 584 (2010).

[26]  S. Xu, Y. Han, X. Chen, Z. Wu, L. Wang, T. Han, W. Ye, H. Lu, G. Long, Y. Wu, J. Lin, Y. Cai, K. M. Ho, Y. He, and N. Wang, Nano Lett. **15**, 2645 (2015).

[27]  P. Gehring, B. F. Gao, M. Burghard, and K. Kern, Nano Lett. **12**, 5137 (2012).

[28]  T. Taniguchi and K. Watanabe, J. Cryst. Growth **303**, 525 (2007).

[29]  C. R. Dean, a F. Young, I. Meric, C. Lee, L. Wang, S. Sorgenfrei, K. Watanabe, T. Taniguchi, P. Kim, K. L. Shepard, and J. Hone, Nat. Nanotechnol. **5**, 722 (2010).

[30]  W. Dang, H. Peng, H. Li, P. Wang, and Z. Liu, Nano Lett. **10**, 2870 (2010).

[31]  A. S. Mayorov, R. V Gorbachev, S. V Morozov, L. Britnell, R. Jalil, L. a Ponomarenko, P. Blake, K. S. Novoselov, K. Watanabe, T. Taniguchi, and a. K. Geim, Nano Lett. **11**, 2396 (2011).

[32]  M. H. D. Guimarães, P. J. Zomer, J. Ingla-Aynés, J. C. Brant, N. Tombros, and B. J. van Wees, Phys. Rev. Lett. **113**, 86602 (2014).

[33]  S. Xu, Z. Wu, H. Lu, Y. Han, G. Long, X. Chen, T. Han, W. Ye, Y. Wu, J. Lin, J. Shen, Y. Cai, Y. He, F. Zhang, R. Lortz, C. Cheng, and N. Wang, 2D Mater. **3**, 21007 (2016).

[34]  X. Chen, Y. Wu, Z. Wu, Y. Han, S. Xu, L. Wang, W. Ye, T. Han, Y. He, Y. Cai, and N. Wang, Nat. Commun. **6**, 7315 (2015).

[35]  K. S. Novoselov, A. Mishchenko, A. Carvalho, and A. H. Castro Neto, Science (80-. ). **353**, aac9439 (2016).

[36]  Z. Fei, T. Palomaki, S. Wu, W. Zhao, X. Cai, B. Sun, P. Nguyen, J. Finney, X. Xu, and D. H. Cobden, Nat Phys **13**, 677 (2017).

[37]  J. I.-J. Wang, Y. Yang, Y.-A. Chen, K. Watanabe, T. Taniguchi, H. O. H. Churchill, and P. Jarillo-Herrero, Nano Lett. **15**, 1898 (2015).

[38]  D. M. Pozar, *Microwave Engineering*, 3rd ed. (Wiley, 2005).

[39]  C. N. Berglund, IEEE Trans. Electron Devices **ED-13**, 701 (1966).

[40]  Y. Xia, D. Qian, D. Hsieh, L. Wray, A. Pal, H. Lin, A. Bansil, D. Grauer, Y. S. Hor, R. J. Cava, and M. Z. Hasan, Nat Phys **5**, 398 (2009).

[41]  M. Orlita, B. a. Piot, G. Martinez, N. K. S. Kumar, C. Faugeras, M. Potemski, C. Michel, E. M. Hankiewicz, T. Brauner, Č. Drašar, S. Schreyeck, S. Grauer, K. Brunner, C. Gould, C. Brüne, and L. W. Molenkamp, Phys. Rev. Lett. **114**, 186401 (2015).

[42]  H. He, B. Li, H. Liu, X. Guo, Z. Wang, M. Xie, and J. Wang, Appl. Phys. Lett. **100**, 32105 (2012).

[43]  L. Wu, M. Brahlek, R. Valdés Aguilar, a. V. Stier, C. M. Morris, Y. Lubashevsky, L. S. Bilbro, N. Bansal, S. Oh, and N. P. Armitage, Nat. Phys. **9**, 410 (2013).

[44]  J. Son, K. Banerjee, M. Brahlek, N. Koirala, S.-K. Lee, J.-H. Ahn, S. Oh, and H. Yang, Appl. Phys. Lett. **103**, 213114 (2013).

[45] L. Veyrat, F. Iacovella, J. Dufouleur, C. Nowka, H. Funke, M. Yang, W. Escoffier, M. Goiran, B. Eichler, O. G. Schmidt, B. Büchner, S. Hampel, and R. Giraud, Nano Lett. **15**, 7503 (2015).

[46] A. A. Taskin, S. Sasaki, K. Segawa, and Y. Ando, Adv. Mater. **24**, 5581 (2012).

[47] H. Steinberg, J.-B. Laloë, V. Fatemi, J. S. Moodera, and P. Jarillo-Herrero, Phys. Rev. B **84**, 233101 (2011).

[48] H. Liu, S. Liu, Y. Yi, H. He, and J. Wang, 2D Mater. **2**, 045002 (2015).

[49] N. Koirala, M. Brahlek, M. Salehi, L. Wu, J. Dai, J. Waugh, T. Nummy, M. G. Han, J. Moon, Y. Zhu, D. Dessau, W. Wu, N. P. Armitage, and S. Oh, Nano Lett. **15**, 8245 (2015).

[50] J. Y. Park, G.-H. Lee, J. Jo, A. K. Cheng, H. Yoon, K. Watanabe, T. Taniguchi, M. Kim, P. Kim, and G.-C. Yi, 2D Mater. **3**, 35029 (2016).

[51] D. Kim, S. Cho, N. P. Butch, P. Syers, K. Kirshenbaum, S. Adam, J. Paglione, and M. S. Fuhrer, Nat. Phys. **8**, 459 (2012).

[52] D. Hsieh, Y. Xia, D. Qian, L. Wray, J. H. Dil, F. Meier, J. Osterwalder, L. Patthey, J. G. Checkelsky, N. P. Ong, A. V Fedorov, H. Lin, A. Bansil, D. Grauer, Y. S. Hor, R. J. Cava, and M. Z. Hasan, Nature **460**, 1101 (2009).

[53] G. L. Yu, R. Jalil, B. Belle, A. S. Mayorov, P. Blake, F. Schedin, S. V Morozov, L. a Ponomarenko, F. Chiappini, S. Wiedmann, U. Zeitler, M. I. Katsnelson, a K. Geim, K. S. Novoselov, and D. C. Elias, Proc. Natl. Acad. Sci. U. S. A. **110**, 3282 (2013).

[54] M. Bianchi, D. Guan, S. Bao, J. Mi, B. B. Iversen, P. D. C. King, and P. Hofmann, Nat. Commun. **1**, 128 (2010).

[55] N. P. Butch, K. Kirshenbaum, P. Syers, A. B. Sushkov, G. S. Jenkins, H. D. Drew, and J. Paglione, Phys. Rev. B **81**, 241301 (2010).

[56] H. Köhler and C. R. Becker, Phys. Status Solidi **61**, 533 (1974).

[57] V. Fatemi, B. Hunt, H. Steinberg, S. L. Eltinge, F. Mahmood, N. P. Butch, K. Watanabe, T. Taniguchi, N. Gedik, R. C. Ashoori, and P. Jarillo-Herrero, Phys. Rev. Lett. **113**, 206801 (2014).

[58] Y. Shao, K. W. Post, J.-S. Wu, S. Dai, A. J. Frenzel, A. R. Richardella, J. S. Lee, N. Samarth, M. M. Fogler, A. V. Balatsky, D. E. Kharzeev, and D. N. Basov, Nano Lett. **17**, 980 (2017).

[59] D. A. Kozlov, D. Bauer, J. Ziegler, R. Fischer, M. L. Savchenko, Z. D. Kvon, N. N. Mikhailov, S. A. Dvoretsky, and D. Weiss, Phys. Rev. Lett. **116**, 166802 (2016).

[60] H. Preier, Appl. Phys. **20**, 189 (1979).

[61] B. A. Assaf, T. Phuphachong, V. V Volobuev, G. Bauer, G. Springholz, L.-A. de Vaulchier, and Y. Guldner, Npj Quantum Mater. **2**, 26 (2017).